\documentclass[aps,final,1p, showpacs,twocolumn]{revtex4}
\usepackage{epsfig}
\usepackage{amssymb}
\usepackage{amsthm}
\usepackage[nodots]{numcompress}

\begin{document}

\title{Numerical Studies on Antiferromagnetic  Skyrmions in   Nanodisks by Means of A New Quantum Simulation Approach }

\author{
Zhaosen Liu$^{a,b}$\footnote{Email: liuzhsnj@yahoo.com},
Hou Ian$^b$\footnote{Email: houian@umac.mo}
}
\address{$^a$Department of Applied  Physics,
 Nanjing University of Information Science and Technology,    Nanjing 210044, China\\
$^b$Institute of Applied Physics and Materials Engineering, FST,
University of Macau, Macau}

%\affiliation{}

\begin{abstract}
We employ a self-consistent simulation approach based on quantum physics   here to study the magnetism of antiferromagnetic   skyrmions  formed on manolayer nanodisk planes. We find that if the  disk  is   small and   the  Dzyaloshinsky-Moriya (DM)   interaction is weak, a single magnetic vortex may be formed on the  disk plane. In such a case,  when  uniaxial  anisotropy   normal to the disk plane is further considered, the magnetic configuration  remains unchanged,   but the magnetization   is enhanced in that direction, and  reduced in other two perpendicular orientations. Very similarly, a  weak external magnetic field normal  to the disk plane cannot obviously affect the spin structure of the nanodisk;   however, when it is sufficiently strong, it can destroy the  AFM skyrmion  completely. On the other hand, by increasing  DM interaction  so that  the disk diameter is a few times larger than the DM length,   more self-organized magnetic domains, such as vortices and strips, will  be  formed in the disk plane.  They   evolve with decreasing temperature,  however always symmetric about a geometric axis of the square unit cell.  We further find that in this case introducing  normal  magnetic anisotropy    gives rise to
the re-construction of  AFM   single-vortex  structure  or skyrmion  on the  disk plane, which provides a way  to create and/or
stabilize such spin texture in   experiment.
\end{abstract}


\maketitle

\section{Introduction}

The concepts of skyrmions were  originally introduced
 by a   particle physicist, Tony Skyrme, to describe the localized, particle-like
structures in the field of pion particles in the early 1960s \cite{Skyrme}. About  30 years later,
 Bogdanov and  Yablonskii theoretically predicted  \cite{Bogdanov} that they  could exist in magnets
 when  a  chiral Dzyaloshinsky-Moriya (DM)   interaction  \cite{Dzy,Moriya,Moriya2} is present. Indeed,
 it was later found in experiments that magnetic skyrmions exist in helical magnets, such as MnSi   and
 Fe$_{1-x}$Co$_x$Si \cite{Munzer,Yu, Tonomura}, and
 DM interaction favors canted spin configuration
  \cite{Muhlbauer, Munzer, Yu, Tonomura, Heinze,Adams,Adams2,Bogdanov2,Bogdanov3,Pfleiderer,Chung}.
Most skyrmions found in helimagnets were induced by an external magnetic field   at low
temperatures \cite{Munzer,Yu, Tonomura,Kiselev}. For example, Heinze et al. \cite{Heinze} observed a spontaneous
 atomic-scale magnetic ground-state skyrmion lattice in a mono-layer Fe film at a low temperature  about 11 K.
 However,   Yu et al. \cite{Yu}   obtained a skyrmion crystal near room-temperature in  FeGe  with a high
  transition temperature (280 K)  by applying a magnetic field.

% In addition, chirality such as that of skyrmion in nanoscale  magnets plays   important role in spintronic
% devices, since the spin instead of  the  charge of an electron is used  for data storage, transmission and manipulation \cite{Bode}.

So far, ferromagnetic (FM) skyrmions have been intensively investigated both theoretically and experimentally.
However,  the DM interaction    is more generally  found in antiferromagnetic (AFM) materials
than ferromagnetic  materials. Most recent experiments on FM skyrmions rely  on the presence
of the interfacial DM interaction to stabilize skyrmions. In contrast,
bulk DM interaction is more prevalent in  antiferromagnets \cite{Dzy,Moriya2}, and they
are  considerably more abundant in nature than ferromagnets.

Skyrmions have  been predicted to appear in the ground state of doped antiferromagnetic   insulators \cite{Kivelson,Sushkov,Ando}.
However,  it is difficult to   identify  these  isolated skyrmions. Neutron scattering, for example, would not be an  effective probe, since these skyrmions  do not form a lattice, whereas their  signatures on transport may be screened by the insulating character of the carriers \cite{Raicvic}. Based on their experimental observation, Rai{\v c}vi\'c et al. concluded that the low-temperature  magnetic and transport properties of the AFM La$_2$Cu$_{1-x}$Li$_x$O$_4$  provided the first experimental support for the presence of skyrmions in AFM insulators.

 Recently, Huang et al. \cite{Huang} simulated   the  creation process of skyrmion in a two-dimensional (2D)  antiferromagnetic system to  investigate the dynamics of the created skyrmions, and  observed   stable skyrmions  even at long time scales.
 So far,  many researchers have done extensive  studies on the static properties of 2D FM skyrmions. Therefore,  it is obviously necessary and  meaningful to investigate   how  the magnetism of the 2D AFM skyrmions are influenced by external magnetic field,     Heisbenerg exchange, anisotropic  and DM interactions, so as to  find   ways to create or stabilize the AFM skyrmion in  experiments.  For the purpose, this work has been done.

In a just finished work \cite{LiuIan1}, we investigated the magnetic and thermodynamic properties of
mono-layer nanodisks with the co-exitance of   DM and FM  Heisenberg interactions
by means of a new quantum simulation approach  we develop  in recent years \cite{liujpcm,liupssb}.
We found there that the chirality of   the single magnetic vortex on a small nanodisk   is only determined
by  the sign of DM interaction parameter, no matter  an external
 magnetic field is absent or applied normal to the disk plane,  however the applied magnetic field
  perpendicular to the disk-plane is able to stabilize the vortex structure and induce skyrmions \cite{Munzer,Yu, Tonomura,Kiselev}.

In the present work, the new quantum simulation approach is applied to AFM  mono-layer  nanodisks
with the co-existence of   Heiseinberg and DM  interaction as well. We find that for small disk of   weak   DM interaction,
single AFM skyrmion is always formed on the disk plane. Further inclusion  of   uniaxial  anisotropy or weak external magnetic field   normal to the disk plane causes no obvious change in the spin configuration, but they do enhance the magnetization
 in that direction and reduce the other two in-plane components. However,  if this  applied magnetic field
 is strong enough,  the in-plane  AFM Skyrmion  will be completely destroyed.   Moreover, by increasing  the  DM interaction,
  more self-organized magnetic domains will appear on the disk plane. They   evolve with varying
temperature, but always symmetric about a geometric axis of the square unit cell.  In this case,  an
 uniaxial  magnetic anisotropy   normal to the disk plane   is able to force the multi-domain structure
  merge to form a  single   AFM  vortex. In another word, the anisotropy can induce and/or stabilize the AFM skyrmion,
  which the experimentalists may be very interested.

\section{Modeling and Computational Algorithm}

The Hamiltonian of this sort of  nanosystems   can be written as
 \cite{Fert1990, Heide2008, Heinze,Kwon13, Thiaville2012,  Fert2013, Sampaio2013, Rohart2013, YMLuo, Gong2015}
\begin{eqnarray}
{\cal H} = & -\frac{1}{2}\sum_{i,j\neq i}\left[{\cal J}_{ij}{
\vec{S}_i \cdot }\vec{S}_j   - D_{ij}{\vec{r}_{ij}\cdot(\vec{S}_i\times}\vec{S}_j)\right]\nonumber\\
\space & -K_A\sum_i\left(\vec{S}_i\cdot \hat{n}  \right)^2  -\mu_Bg_S{\vec B }\cdot\sum_i{\vec S}_i     \;,
\label{hamil}
\end{eqnarray}
where the first and second terms represents the Heisenberg
exchange and  DM interactions with strength of
${\cal J}_{ij}$ and $D_{ij}$ between every pair of neighboring
spins  sitting at the $i$- and $j-$th sites,  respectively, the third
term denotes the uniaxial anisotropy along $\hat{n}$,
assumed   to be normal to the disk plane here, and the
last one is the Zeeman energy of the system   within external magnetic  field ${\vec B}$. For simplicity, we consider  in the
current work a round mono-layer  nanodisk consisting of
$S$ = 1 spins which interact antiferromagnetically only with their nearest neighbors
uniformly, that is, ${\cal J}_{ij}$ = ${\cal J}$ and
$D_{ij} = D$, across the whole disk plane.
In our model, the spins are quantum operators instead of the classical vectors.
Since $S = 1$, the matrices of the three spin components   are given by
{\footnotesize
\begin{eqnarray}
S_x = &\frac{1}{2}\left(
\begin{array}{ccc}
0 &  \sqrt{2}  & 0\\
\sqrt{2} & 0 & \sqrt{2}  \\
0 & \sqrt{2}  & 0\\
\end{array}
\right)\;, & S_y =\frac{1}{2i}\left(
\begin{array}{ccc}
0 &  \sqrt{2}  & 0\\
-\sqrt{2} & 0 & -\sqrt{2}  \\
0 & \sqrt{2}  & 0\\
\end{array}
\right)\;\\ \nonumber      S_z = & \left(
\begin{array}{ccc}
1 &  0  & 0\\
0 & 0 & 0 \\
0 & 0  & -1\\
\end{array}
\right) & \space \\ \nonumber\;
\end{eqnarray}\nonumber
}
respectively.

Our simulation approach, which is facilitated by a self-consistent
algorithm,  so  called  as the SCA approach,  was assumed to be
based on the principle of the least  (free)  energy
\cite{liujpcm,liupssb}.   Thus,  as a computational code implemented with this   algorithm runs,
 all magnetic moments in the sample are rotated and their
magnitudes adjusted by the local effective magnetic field to
minimize the total (free) energy of the whole nanosystem
spontaneously according to the law of least (free) energy, so that
the code can finally converge down to the equilibrium state
automatically without the need to minimize
the total (free) energy elaborately in every simulation step.

%%%%%%%%%Fig1%%%%%%%%%%%%%%%
\begin{figure*}[htb]
 \centerline{\epsfig{file=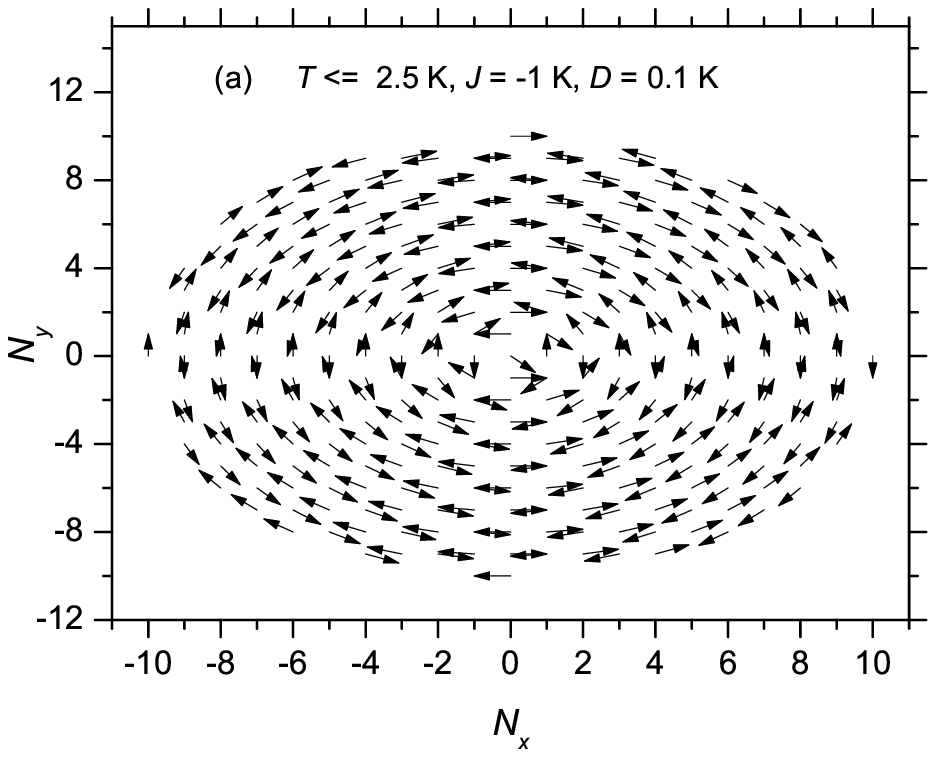,width=.42\textwidth,height=6.2cm,clip=}
 \epsfig{file=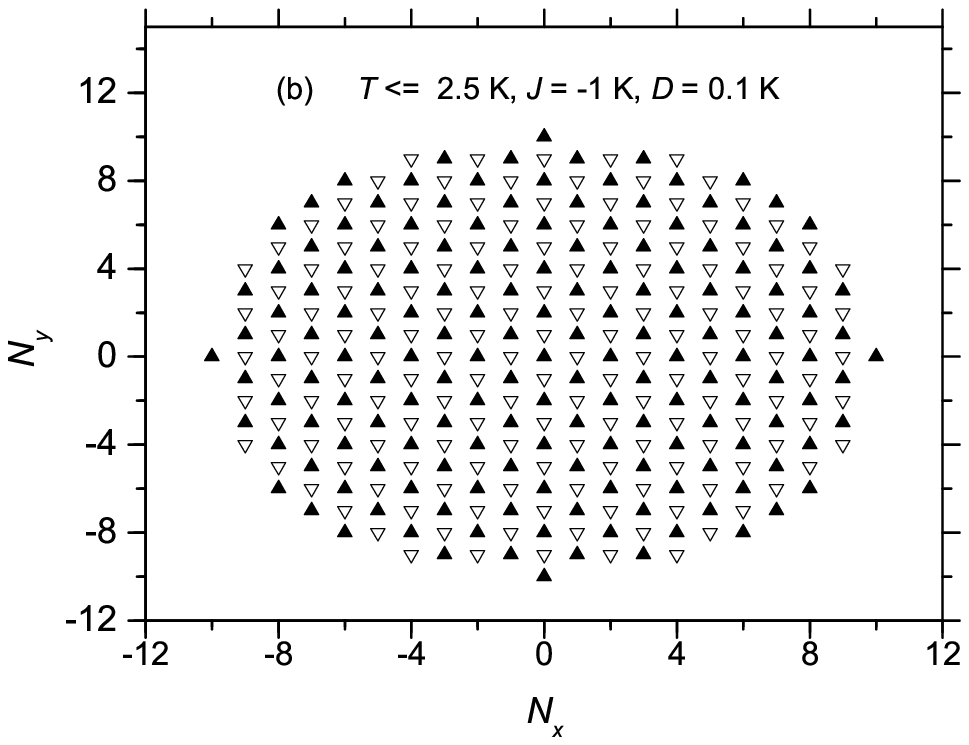,width=0.40\textwidth,height=6.2cm,clip=}}
\begin{center}
\parbox{14cm}{\small{{\bf Figure 1.}
 The    (a) $xy$, and (b) $z$ components of calculated spin configurations projected onto the  nanodisk plane.
  Here ${\cal J} $ = -1 K, $D$ =0.1 K, and $R = 10a$, respectively. }}
\end{center}
\end{figure*}
%%%%%%%%%%%%%%%%%%%%%%%

All of our recent simulations     are  started from a random
magnetic configuration  and from a temperature well   above  the magnetic transition
temperature $T_M$, then   carried out stepwise down to very low
temperatures with an iteration step $\Delta T < 0$.
At any temperature, if the difference $(|\langle \vec{S}'_i\rangle -
\langle \vec{S}_i\rangle|)/|\langle \vec{S}_i\rangle|$ between two successive iterations for
every spin is less than a very small given value $\tau_0$,
convergency is believed to be reached.

\section{Calculated Results}

\subsection{Simulations for Nanodisk without Uniaxial Anisotropy and External Magnetic Field}

%%%%%%%%Fig2%%%%%%%%%%%%%%%
\begin{figure*}[ht]
\centerline{
 \epsfig{file=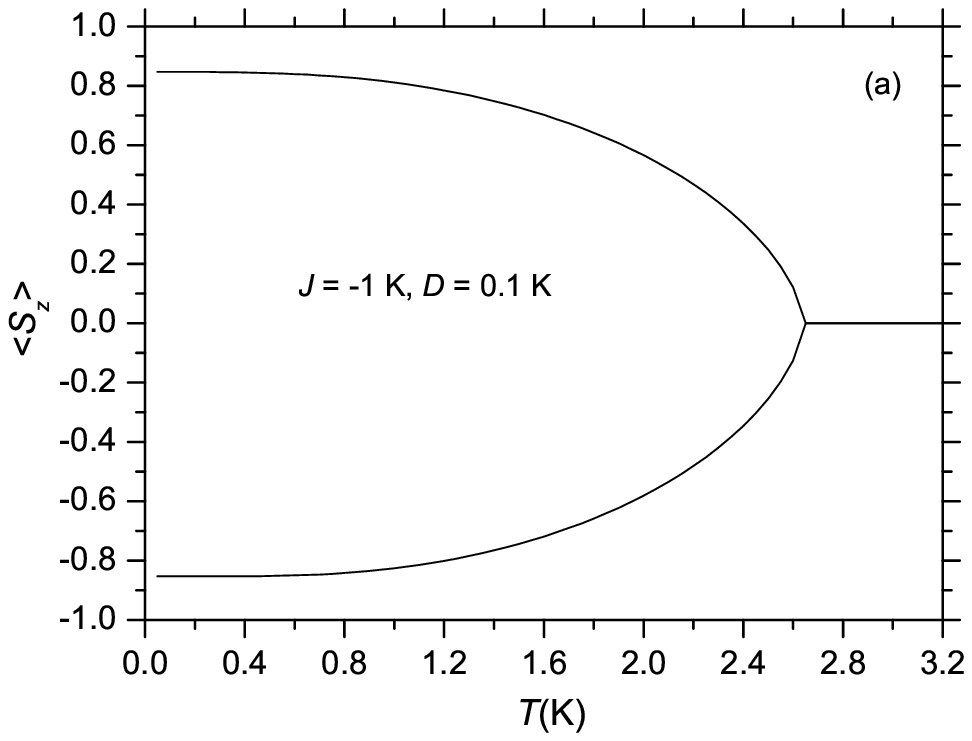,width=0.40\textwidth,height=6.2cm,clip=}
\epsfig{file=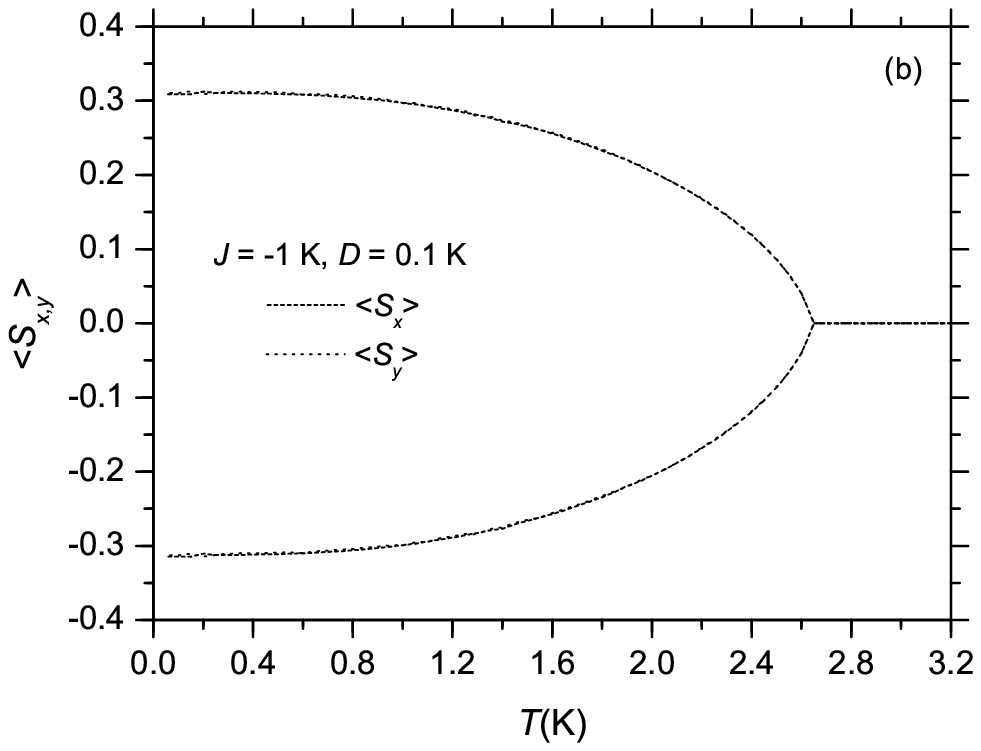,width=0.40\textwidth,height=6.2cm,clip=} }
\centerline{
 \epsfig{file=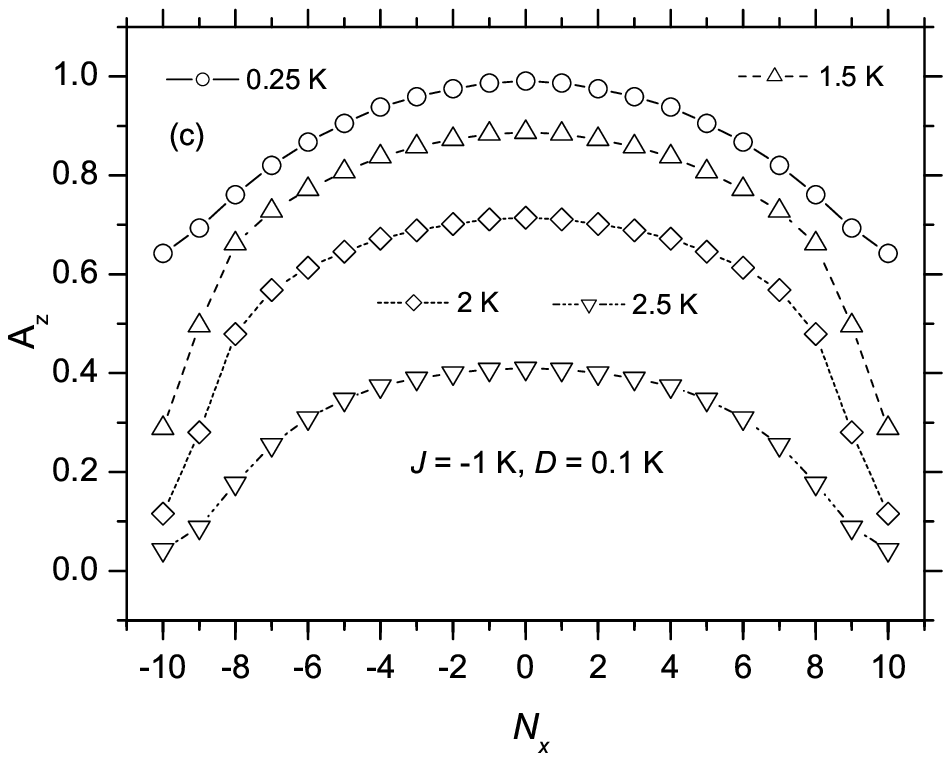,width=0.40\textwidth,height=6.2cm,clip=}
\epsfig{file=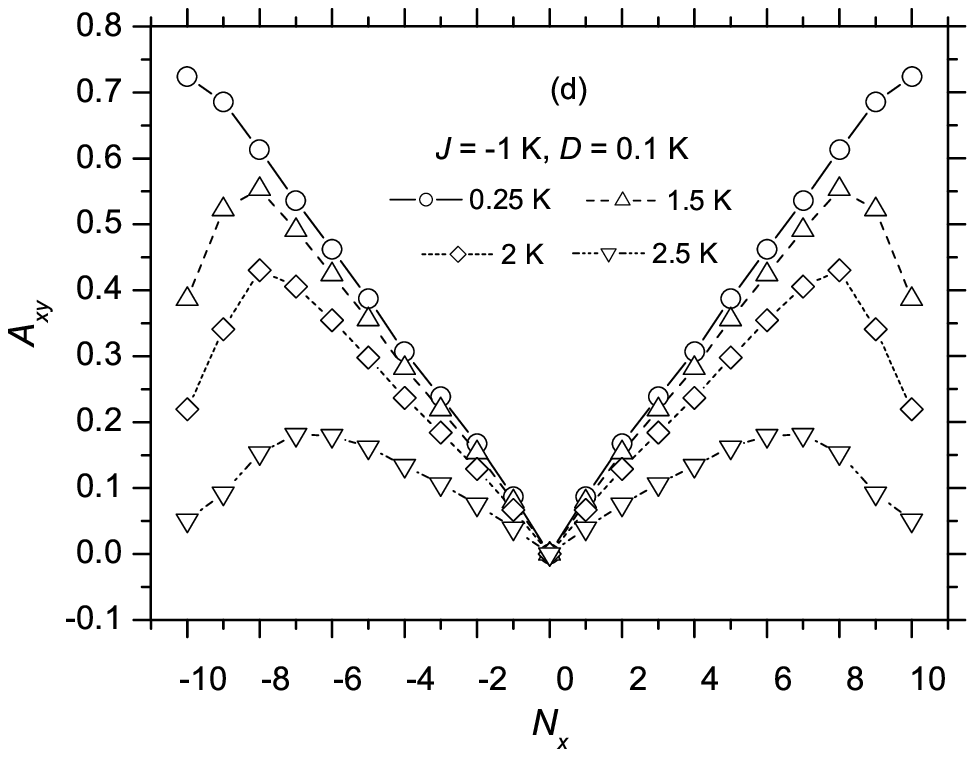,width=0.40\textwidth,height=6.2cm,clip=} }
\begin{center}
 \parbox{14cm}
 {\small{{\bf Figure 2.}
 Calculated spontaneous (a)   $\langle S_z\rangle $, and (b) $\langle S_{x,y}\rangle $ for
the mono-layer AFM nanodisk as functions of temperature;    (c)  $A_z$, and (d) $A_{xy}$
 as functions of the distances from the center of the nanodisk at four different temperatures.
  Here $R$ = 10a, ${\cal J}$ = -1 K, and $D$ = 0.1 K, respectively.   }}
 \end{center}
\end{figure*}
%%%%%%%%%%%%%%%%%%%%%%%%%%%%%%%%%%%%%%
To investigate the effects of DM interaction,  the magnetic anisotropy was neglected in simulations  at the beginning. And
 to visualize the spin configuration   clearly, we    considered   a very tiny   round mono-layer nanodisk,
 its radius $R$ = 10$a$,   where $a$ is the side length of the square crystal unit cell, and   the spins on the disks are
assumed to be   antiferromagnetically   coupled uniformly.
 We   performed simulations with the SCA approach    by assigning  ${\cal J}$ to  -1 K  and $D$  to     0.1 K, respectively.
 To avoid misunderstanding, we  indicate here that   all  parameters used in this paper are scaled with Boltzmann constant $k_B$.

Under the  DM interaction,  magnetic vortex  is formed on the nanodisk, and owing to the antiferromagnetic
Heisenberg interaction, each pair of neighboring spins order oppositely both in-plane and out-plane  below the
transition temperature $T_M \approx$ 2.65 K  as shown in Figure 1(a,b).

Figure 2(a,b) display  our calculated  thermally averaged $\langle S_z\rangle $,    $\langle S_{x}\rangle $ and $\langle
S_{y}\rangle $ for the  nanodisk in the absence of external magnetic field.  The  DM interaction  has
  induced  out-plane magnetic moments \cite{Shinjo,Wachowiak,Choe, Im}, which is much stronger
than the other two components at all temperatures. The three components decay monotonously with
increasing temperature until the transition point $T_M \approx $ 2.65 K, and   the saturated value of $\langle S_z\rangle$
 is approximately 0.85 at very low temperatures, much less than the maximum value $S$ =  1.

%%%%%%%%%Fig3%%%%%%%%%%%%%%%
\begin{figure*}[htb]
\centerline{
\epsfig{file=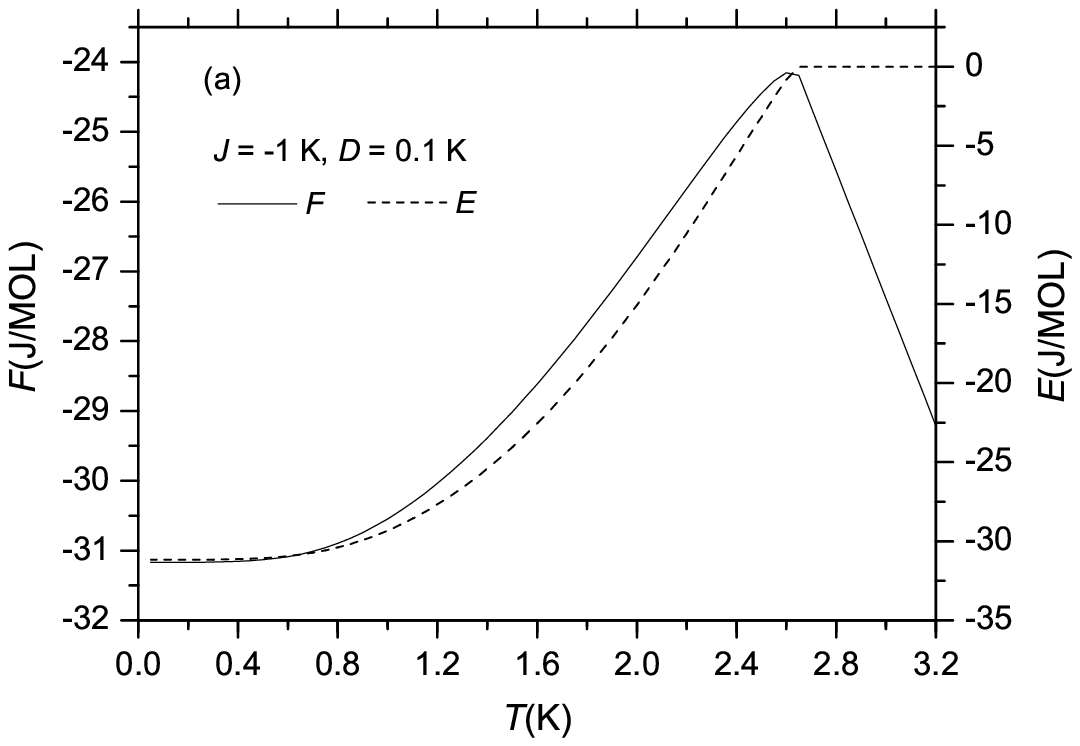,width=0.40\textwidth,height=6.2cm,clip=}
\epsfig{file=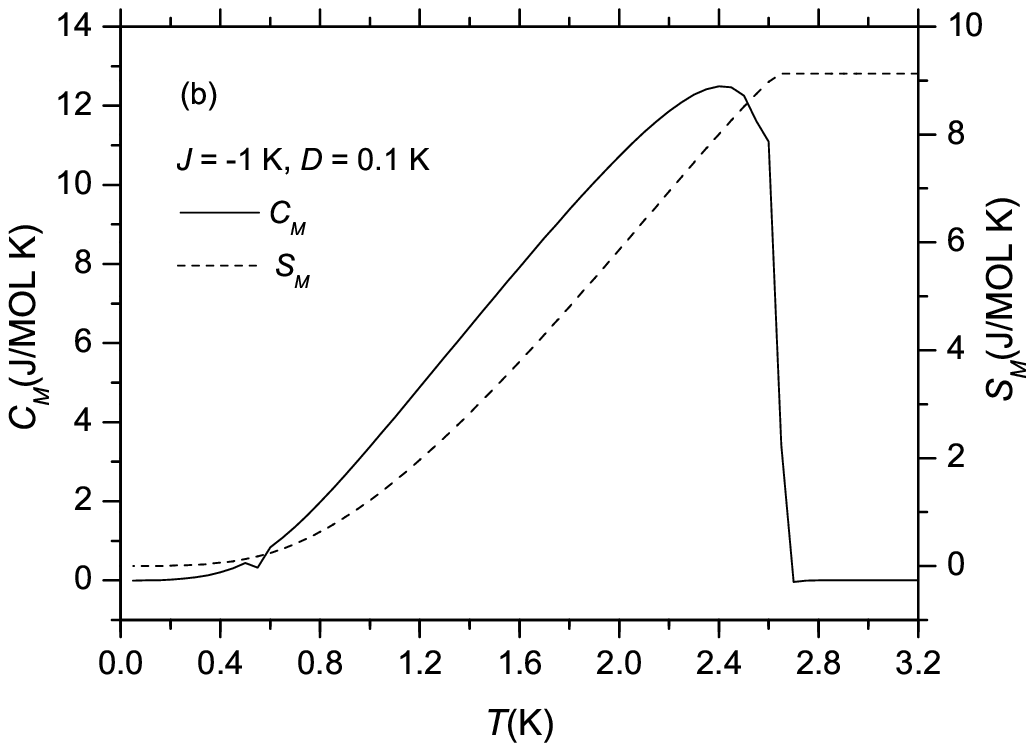,width=0.40\textwidth,height=6.2cm,clip=}}
\begin{center}
\parbox{14cm}{\small{{\bf Figure 3.} Calculated (a) free energy
and energy, (b) magnetic entropy and spefic heat  per mole of the
spins.  Here $R$ = 10a, ${\cal J}$ = -1 K, and $D$ = 0.1 K, respectively.}}
\end{center}
\end{figure*}
%%%%%%%%%%%%%%%%%%%%%%%%%%%%%%%%%%%%%%

To describe the detailed spin configuration on the nanodisk, two new quantities are introduced and defined as
$A_z =  |\langle S_z\rangle|/N_c$  and $A_{xy}   =   \sqrt{\langle S_x\rangle^2+\langle S_y\rangle^2}/N_c$   for the
 out-   and in-plane    components, respectively. Here $N_c(r)$ is the spin number  on the circle
 of radius $r$  around the disk center.   Figure 2(c,d)  display their  variations with changing $r = |N_x|$ at four different
 temperatures. Naturally, the larger the radius, the more spins on the circle. In the inner region   of the disk,
 $A_{xy}$  is   very weaker than $A_z$. That is, therein    the spin are mainly oriented antiferromagnetically out of the
plane, but slightly canted from the normal.    Until $r < $ 6a,   while the radius increases,
 $A_z$ decreases  but $A_{xy}$ grows gradually. That is, as $r$ increases the spins are rotated by the effective magnetic
  field towards the plane,  so that finally,   within the marginal region of the disk the magnitudes of
  $A_z$ and $A_{xy}$ become comparable.

The total    free energy $F$, total   energy $E$, magnetic entropy $S_M$ and specific
heat $C_M$ of this sort of canonical magnetic  systems can be calculated with following formulas
\begin{eqnarray}
 F = -k_BT\log Z_N,\space\space & E = -\frac{\partial}{\partial\beta}\log Z_N\;,\nonumber\\
 S_M = \frac{E}{T}+ k_B\log Z_N, &
 C_M= T\left(\frac{\partial S_M}{\partial T}\right)_B\;,
\end{eqnarray}
successively,  where $\beta =1/(k_BT)$ and $Z_N$ is the partition function of the whole system.  Figure 3(a,b) display  the $F$,
$E$, $S_M$ and $C_M$ curves obtained by means of the SCA  approach for the AFM   nanodisk.
The slopes of $F$, $E$ and $S_M$  curves  change suddenly near $T_M$, which are the signs of phase transition. However, the $C_M$ curve now varies
smoothly    around $T_M$,   in contrast to the sharp peaks observed in the $C_M$ curves near  $T_M$'s of  bulk magnets.   This fact   suggests  that the phase transition behavior of the nanosystem  has been strongly modified  by its  finite size  and  the   spiral DM interaction.

%%%%%%%%%Fig4%%%%%%%%%%%%%%%
\begin{figure*}[htb]
\centerline{
\epsfig{file=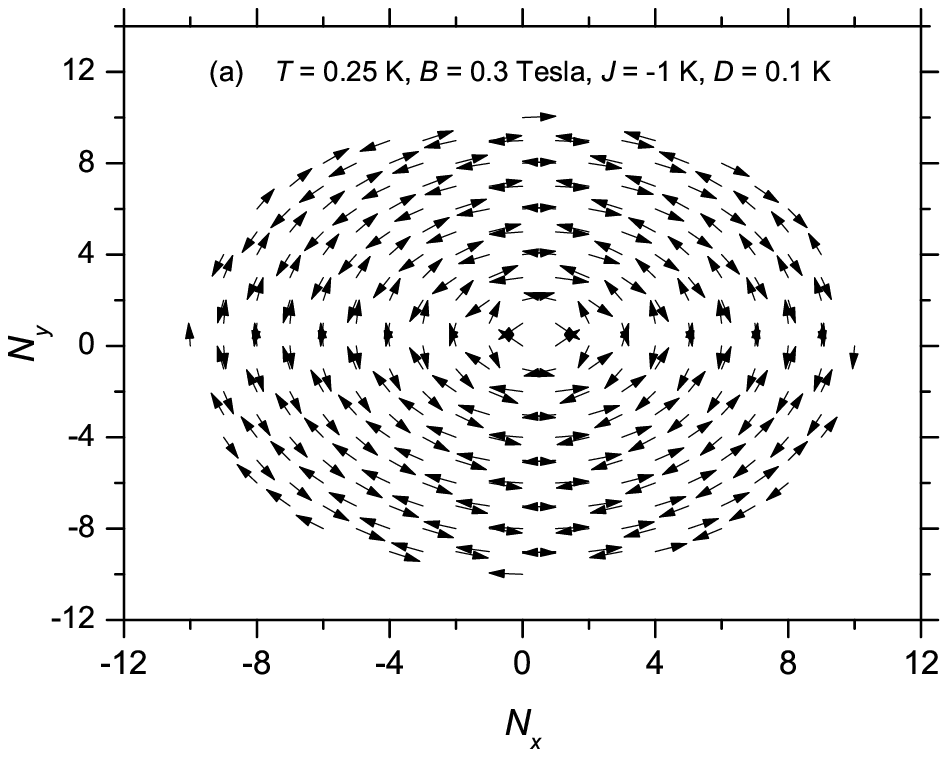,width=0.40\textwidth,height=6.2cm,clip=}
\epsfig{file=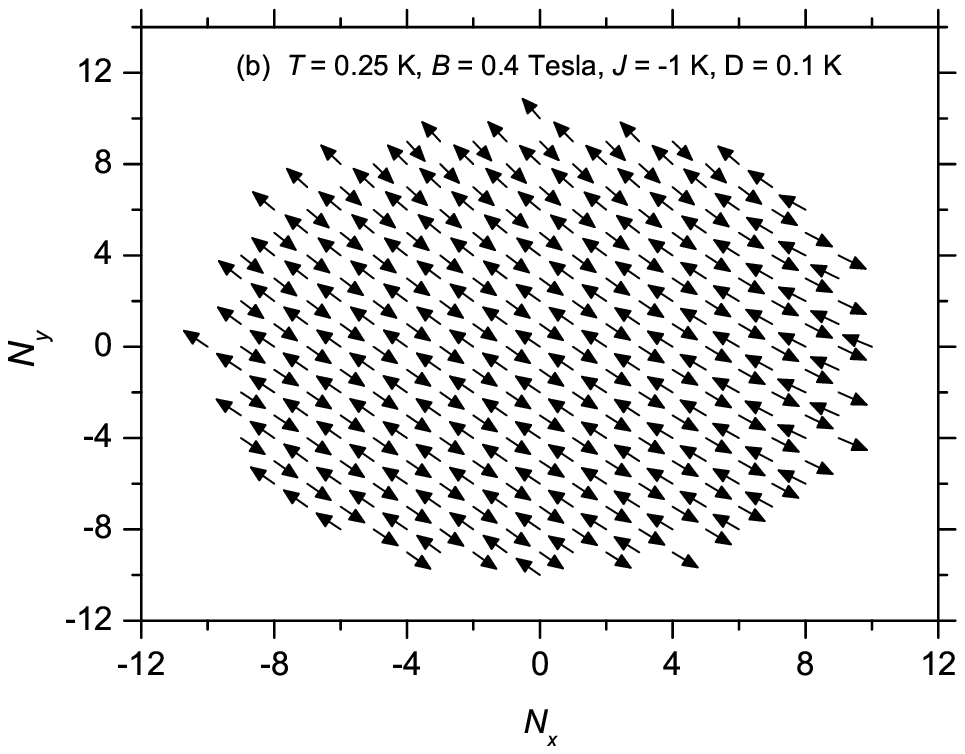,width=0.40\textwidth,height=6.2cm,clip=} }
\begin{center}
\parbox{14cm}{\small{{\bf Figure 4.}  Calculated spin configurations projected onto the disk plane
at $T$ = 0.25 K,  when  an external magnetic field   (a) $B_z$ = 0.3 Tesla,  and (b) $B_z$ = 0.4 Tesla,  is applied normal to the
disk plane. Here ${\cal J}$ = -1 K,   $D$  =  0.1 K and $R = 10a$, respectively.}}
\end{center}
\end{figure*}

\subsection{Effect of External Magnetic Field}
In our previous simulations for FM nanodisks with DM interaction \cite{LiuIan1}, we found that an applied  magnetic field is able
to induce or stabilize the in-plane vortical spin structures,  so that the chiral configuration can maintain to a temperature  well above
$T_M$ as observed in experiments \cite{Munzer,Yu, Tonomura,Kiselev}. So we naturally wonder if this  rule still holds true in the case of  AFM nanodisks.

 For the purpose, by assuming  a  magnetic field   exerted normal to the disk plane  at $T$ = 0.25 K, we did simulations
  for the AFM nanodisk  with   the spin structure  calculated at that temperature in the absence of external magnetic field   as
  the input data. In this circumstance, the nanosystem is able to sustain the external influence to maintain the vortical    structure until $B_z$ = 0.3 Tesla as shown in Fig.4(a). However, when $B_z$ is further increased to 0.4 Tesla, the spiral structure is thus completely
   overcome, whereas the in-plane components of the spins still order antiferromagnetically in the [-1,1,0]
   direction as depicted in Fig.4(b).  This behavior of the nanosystem is easy to understand.  When the external magnetic field along the $z$  direction    is strong enough, the spins  are unable to  align antiferromagnetically in the $z $ direction   any longer, as a result, the in-plane components cannot    form antiferromagnetic   vortices either.  That is,  the formation of an FM (AFM) vortex in the disk plane  depends strongly   on the presence of a spatial region    wherein  the spins    order ferromagnetically (antiferomagnetically) in the normal direction.

\subsection{Effect of Uniaxial Anisotropy}

So far, we have not considered the influence of  magnetic   anisotropy. To study its effects,  it is now assumed
 to be along the $z$-direction to do further simulations.  Since other parameters  are kept unchanged, as expected,
below $T_M$,  $\langle S_z\rangle$ has been enhanced by the anisotropy, but both   $\langle S_x\rangle$ and  $\langle S_y\rangle$ are suppressed for the same reason,  so that the  maximum of $|\langle S_z\rangle|$ is now increased to 0.984, but that of
 $|\langle S_{x,y}\rangle|$   reduced to 0.100 as seen in Fig.5(a).  In addition, all these curves changes gradually, though not smoothly   due to relatively   weak DM interaction,  and  a single magnetic vortex is found on the nanodisk, which prevails in the whole  magnetic phase as depicted in Figure 5(b).

%%%%%%%%Fig5%%%%%%%%%%%%%%%
\begin{figure*}[htb]
\centerline{
 \epsfig{file=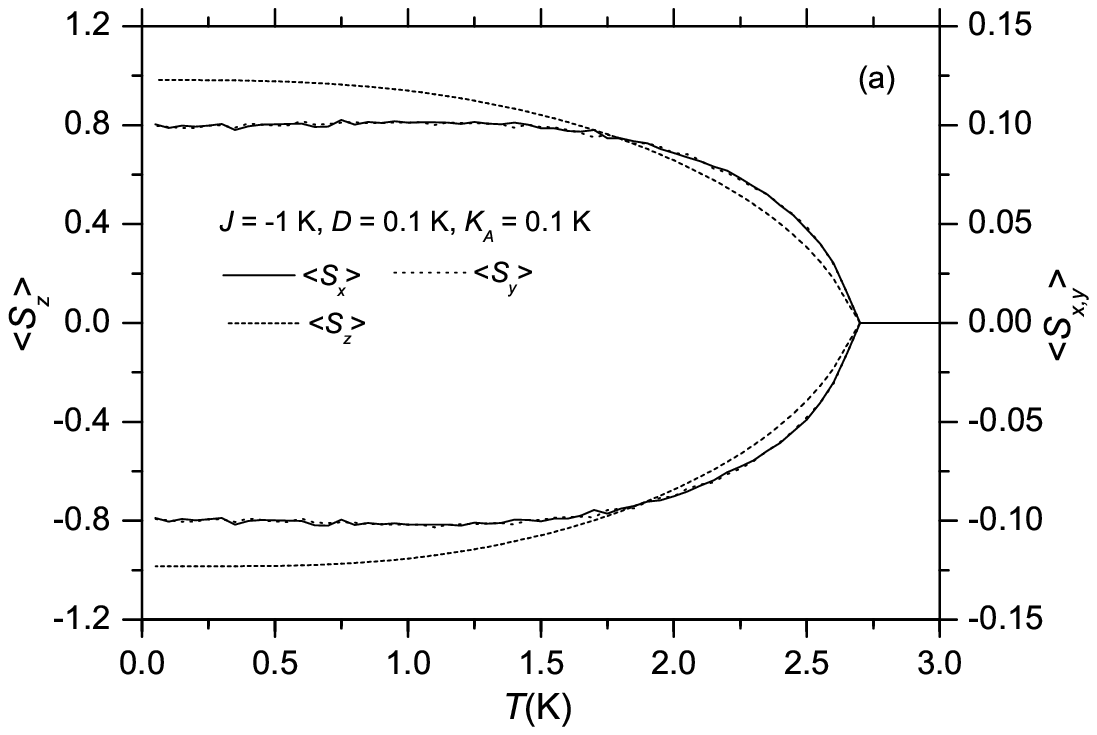,width=0.40\textwidth,height=6.2cm,clip=}
\epsfig{file=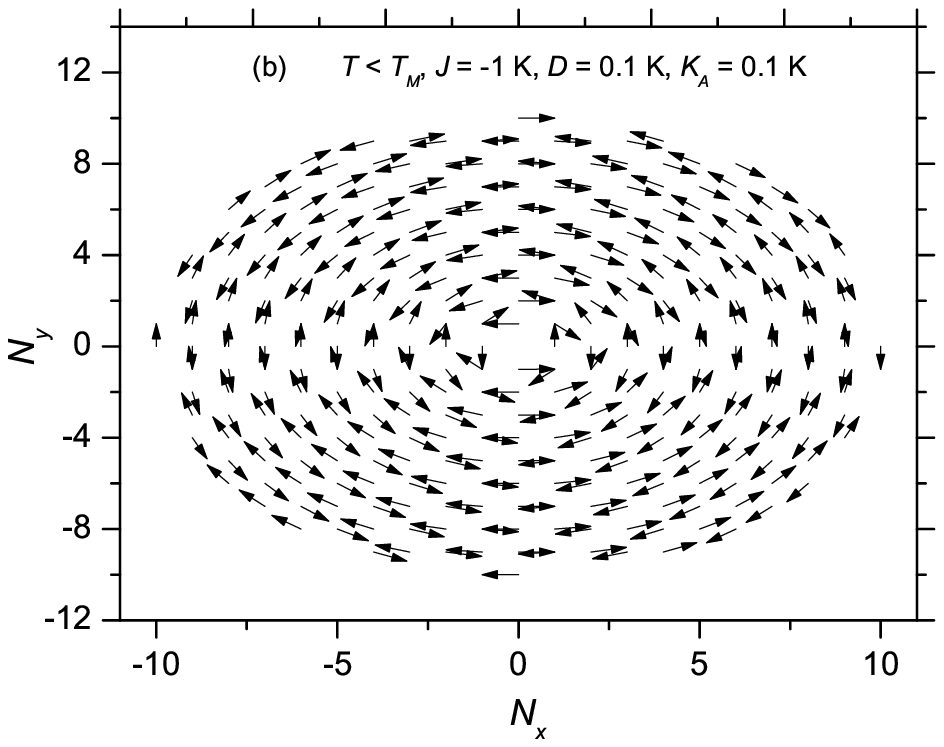,width=0.40\textwidth,height=6.2cm,clip=}}
\begin{center}
 \parbox{14cm}
 {\small{{\bf Figure 5.}   Calculated (a) spontaneous     magnetization for
the mono-layer AFM nanodisk as the functions of temperature, and (b) spin  configurations projected onto the
 $xy$-plane in the temperature region below $T_M$.   Here
$R$ = 10a, ${\cal J}$ = -1 K,   $D$ = 0.1 K and $K_A$ = 0.1 K, respectively.}}
 \end{center}
\end{figure*}
%%%%%%%%%%%%%%%%%%%%%%%%%%%%%%%%%%%%%%

\subsection{Effect of DM Interaction Strength}

To describe the  multi-domain  structures,  a new quantity named  DM length
 has been introduced and defined  as  $\zeta = {\cal J}/D$  which is related to the size of self-organized
  structures, where the distance between two unit grids is  defined as the unity  \cite{Kwon13}. When the
  Monte carlo method is employed,  each grid contains $ n\times n$ atomic sites. We adopt this theory
   by replacing the grid with a spin, and    will see how the theory   works. Thus, as  the disk scale is
   a few times lager than $\zeta$ in the unit
   of lattice parameter $a$, more magnetic structures,  such as strips and vortices, will be formed in the disk plane.
This condition  can realized by either increasing   DM interaction or the lattice size.

To test the idea, we then carried out   simulations for the nanodisk  by increasing the DM interaction to
 $D$ = 0.3 K, but keeping other parameters unchanged.   The calculated $\langle S_z\rangle $,
 $\langle S_{x}\rangle $ and $\langle S_{y}\rangle $ in the absence of external magnetic field
are plotted in Figure 6.   These  magnetization curves are not smooth in the whole low temperature range,
reflecting the  fierce competition between the Heisenberg and DM interactions. The sudden changes  appearing around $T\approx$ 1.25 K
and 0.7 K especially in the $\langle S_{x}\rangle $ and $\langle S_{y}\rangle $ curves  indicate that  phase transitions
   happen  nearby, leading to formations of different magnetic structures.   According to the theory just described,
   now $\zeta = {\cal J}/D \approx 3.333 $, and $2R > \zeta$, so it is expected more self-organized magnetic domains
    will appear in the low temperature region. Above 1.4 K, a single magnetic vortex, as shown in Figure 7(a), occupies the whole disk.  However,
 below $T = $ 1.2 K, a few  magnetic domains appear. The spin configuration   evolves   with decreasing temperature
  until   $T = $ 0.6  K, where  we   observe a very symmetric magnetic structure:  a vertical strip appearing exactly
  in the middle of  two AFM vortices, and this pattern  remains unchanged  down to very low temperature, as displayed in Figure 7(b).
  The two vortex centers are approximately 11$a$ apart, three self-organized domains are involved between, thus the averaged distance of two neighboring  structures is about 3.67$a$,  slightly larger than    $\zeta$ due to the influence of the disk boundary. Therefore, our simulated results   agree   well with  the adopted gird theory.

%%%%%%%%Fig6%%%%%%%%%%%%%%%
\begin{figure*}[htb]
\centerline{
 \epsfig{file=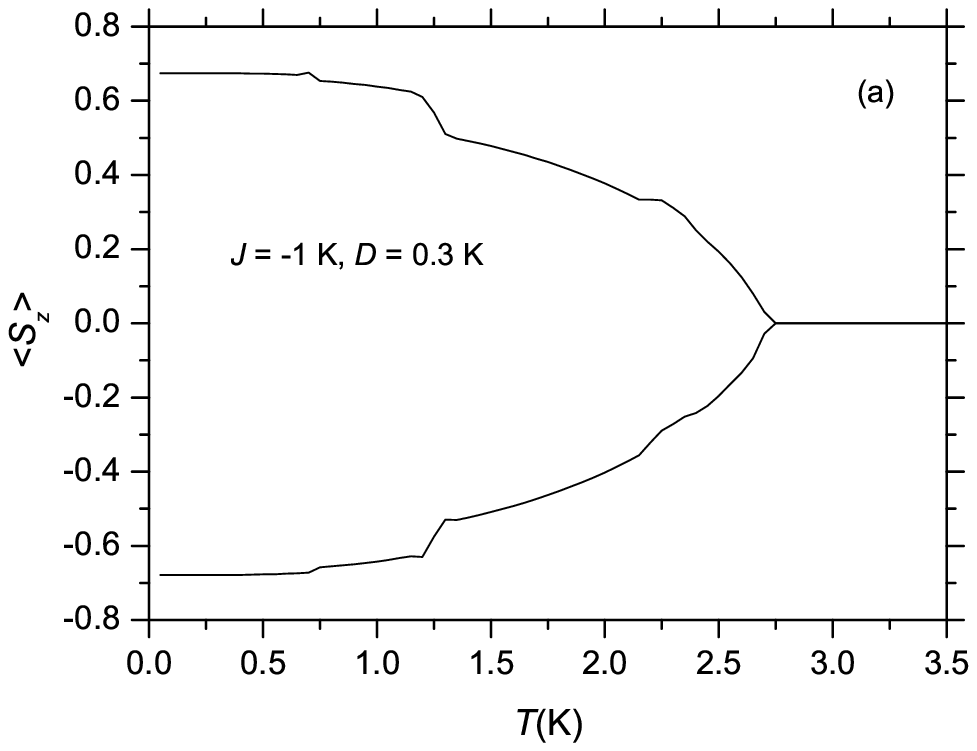,width=0.40\textwidth,height=6.2cm,clip=}
\epsfig{file=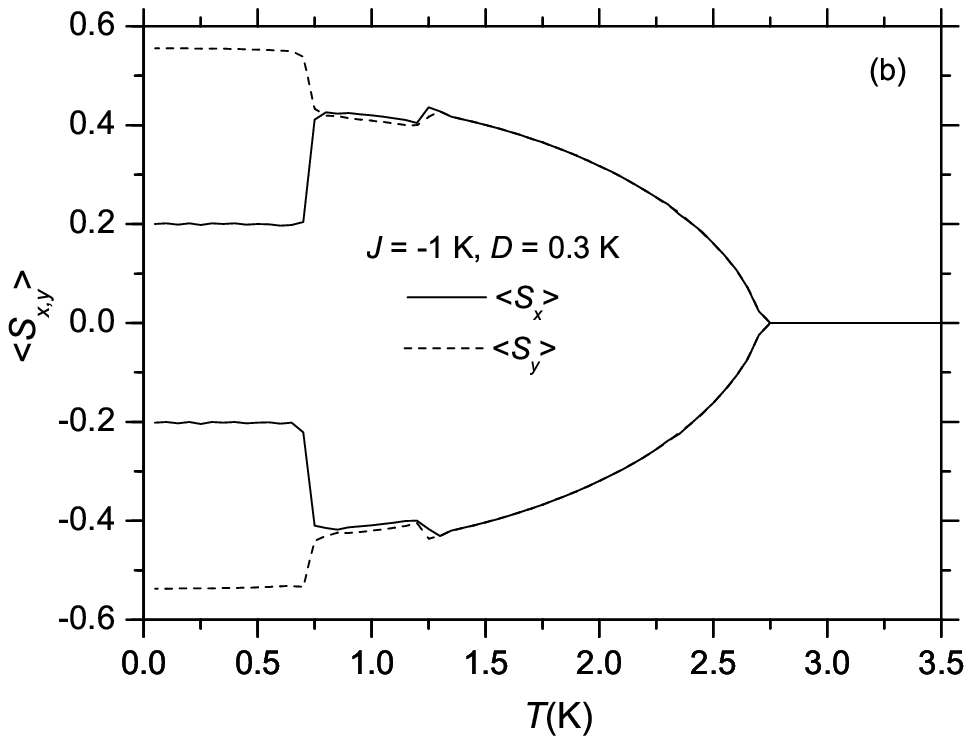,width=0.40\textwidth,height=6.2cm,clip=}}
\begin{center}
 \parbox{14cm}
 {\small{{\bf Figure 6.}  Calculated spontaneous (a)  $\langle S_z\rangle$, and (b) $\langle S_{x,y}\rangle $ for
the mono-layer antiferromagnetic nanoodisk as the functions of temperature. Here $R$ = 10a, ${\cal J}$ = -1 K, and $D$ =
0.3 K, respectively. }}
 \end{center}
\end{figure*}
%%%%%%%%%%%%%%%%%%%%%%%%%%%%%%%%%%%%%%

%%%%%%%%Fig7%%%%%%%%%%%%%%%
\begin{figure*}[htb]
\centerline{
 \epsfig{file=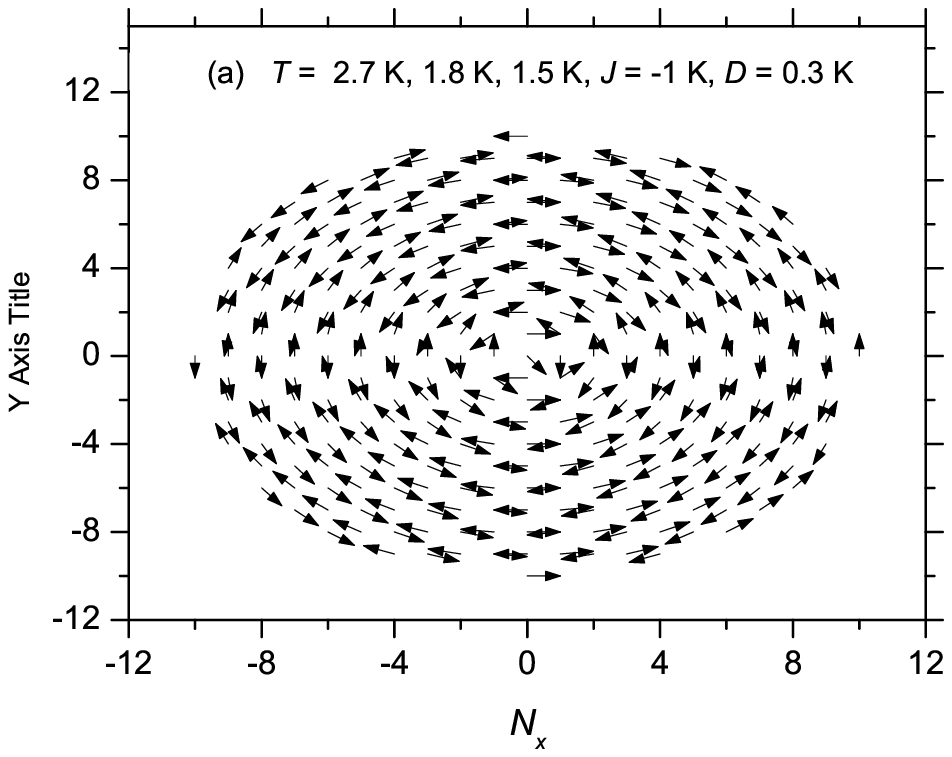,width=0.40\textwidth,height=6.2cm,clip=}
\epsfig{file=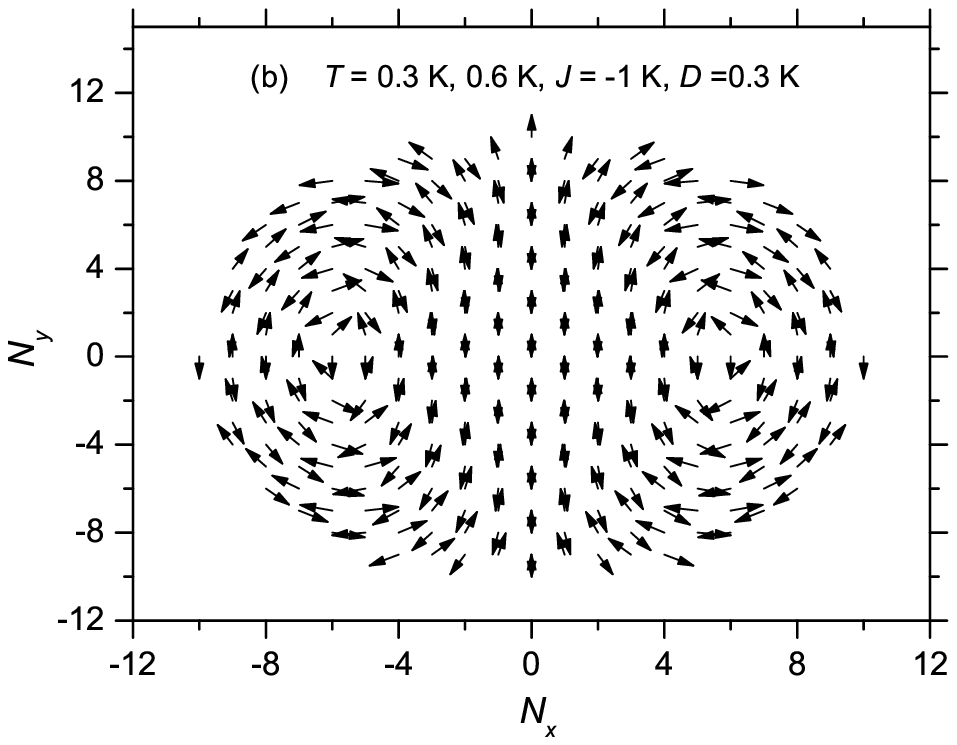,width=0.40\textwidth,height=6.2cm,clip=} }
\begin{center}
 \parbox{14cm}
 {\small{{\bf Figure 7.} Spin  configurations projected onto the $xy$-plane
   calculated  at (a)  $T$ = 2.7, 1.8, 1.5 K, and (b) $T$ = 0.6, 0.3 K.  Here $R$ = 10a, ${\cal J}$ = -1 K, and $D$ =
0.3 K, respectively.  }}
 \end{center}
\end{figure*}
%%%%%%%%%%%%%%%%%%%%%%%%%%%%%%%%%%%%%%

\subsection{Joint Effects of Uniaxial Anisotropy and Strong DM Interaction}

As  described above, a strong DM interaction  usually leads to a multi-domain structure on the disk plane.
On the other hand,  as described above,  the formation of an in-plane AFM skyrmion requires the spins
to  order also  antiferromagnetically in the normal direction, and the  magnetic  anisotropy perpendicular to the disk-plane
 has such an effect. Therefore, we expect that  when a strong DM interaction is present in the nanosystem, which gives rise to multi-domain configuration,   introducing   the   uniaxial anisotropy normal to the disk plane will   recover the single vortex structure.

To test this idea, we carried out simulations by using the  parameters given in Figure 7,  but increasing the
anisotropy strength $K_A$ from zero to 0.1 K. This     anisotropic interaction effectively suppresses
  the  strong DM interaction, consequently, the three components of the magnetization change smoothly and fade gradually
  with increasing temperature below $T_N$   as shown in Figure 8(a), foretelling the appearance  of a single magnetic
    vortex on the disk plane. This prediction is  confirmed by the spin structure that is  stable in the whole magnetic phase,
    as  displayed in Figure 8(b).

%%%%%%%%Fig8%%%%%%%%%%%%%%%
\begin{figure*}[htb]
\centerline{
 \epsfig{file=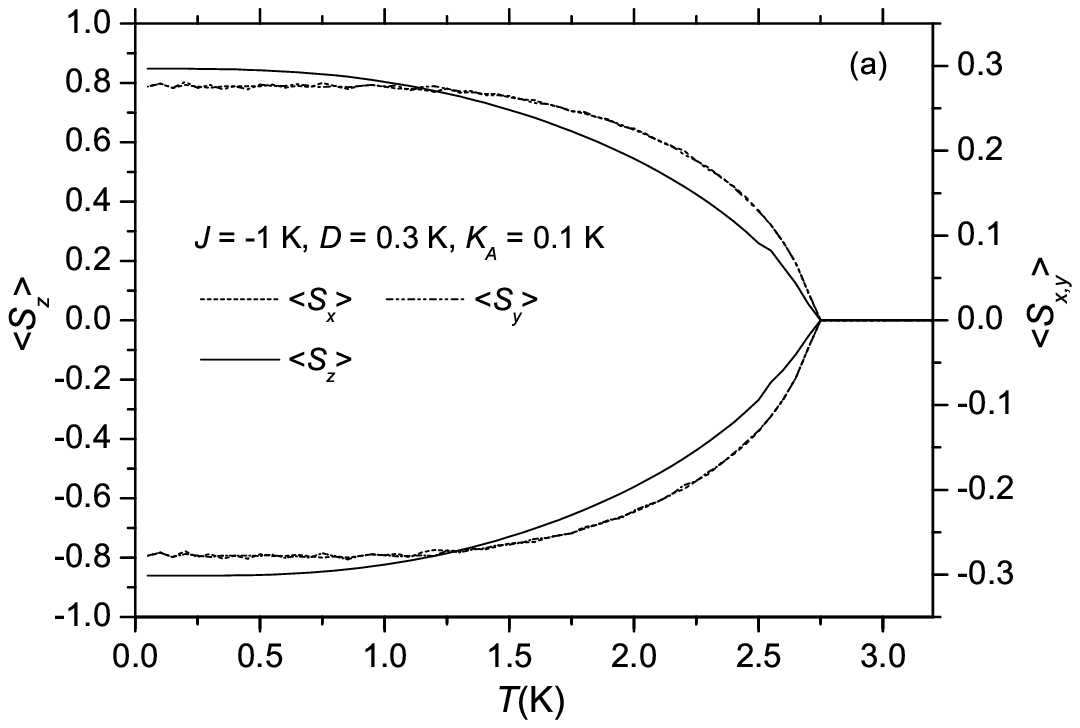,width=0.40\textwidth,height=6.2cm,clip=}
\epsfig{file=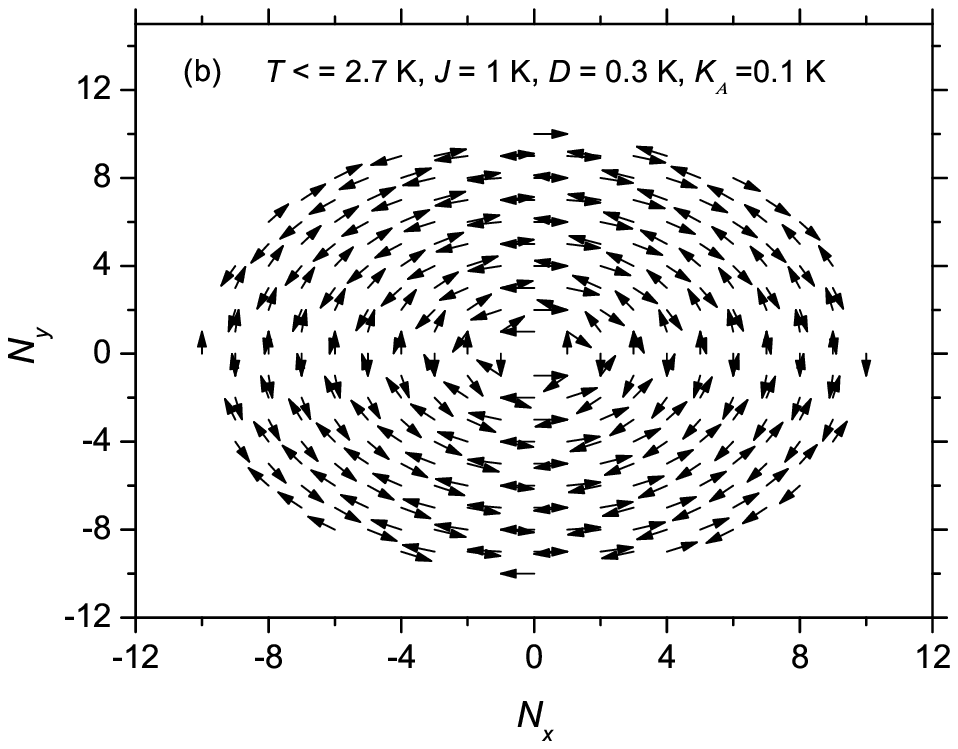,width=0.40\textwidth,height=6.2cm,clip=}}
\begin{center}
 \parbox{14cm}
 {\small{{\bf Figure 8.}
  Calculated (a) magnetization for the mono-layer AFM nanodisk as the functions of temperature,
 (b) spin  configurations projected onto the  $xy$-plane in the  temperature region below $T_M$,
 in the absence of external magnetic field.
 Here $R$ = 10a, ${\cal J}$ = -1 K,  $D$ = 0.3 K and $K_A$ = 0.1 K, respectively.}}
 \end{center}
\end{figure*}
%%%%%%%%%%%%%%%%%%%%%%%%%%%%%%%%%%%%%%

\section{Conclusions and Discussion}

We have successfully carried out simulations for   AFM skyrmions on  manolayer nanodisks  by means of a new quantum
  computational approach. We find that if the disk size is small and  the DM interaction   weak,
single magnetic vortex, or a skyrmion is formed on the
 disk plane. The uniaxial magnetic anisotropy normal to the disk plane does not affect this single
magnetic texture evidently, it can only enhance the magnetic moments in that direction, but reduce the other two
in-plane components. A  weak external magnetic field applied normal to the disk
plane produces similar effects; however, if it is sufficiently strong, it will completely destroy the
magnetic vortex. By increasing the DM interaction
strength so that the disk diameter is a few times larger than the DM length, more self-organized
domains, such as vortices and strips, can   be formed on the disk plane. The spin configuration   evolves with varying
temperature, but is always symmetric about a geometric axis of the square unit cell.
  In this case,   a moderate   uniaxial  magnetic anisotropy   normal to the disk-plane
is able to   suppress  the DM interaction, so that  the multi-domains  merges to  a single  AFM   skyrmion that occupies  the whole  disk plane below
   the transition temperature.

  We have adopted a gird theory  \cite{Kwon13}  to  describe the multi-domain structures on the nanodisks.   The sizes of the magnetic domains and the average distance  between a pair of them  agree  approximately with this modified theory,  as already achieved in  our recent simulations for   FM nanodisks \cite{LiuIan1}.

 We would like to stress finally that  our simulation approach is  based on quantum physics -- the spins appearing in the Hamiltonian are treated as  quantum operators instead of classical vectors, the thermal expectation values of all physical quantities are calculated with quantum formulas. Consequently,
  the computational program is able to run self-consistently, and quickly converge down to equilibrium  state of the magnetic system automatically. Frequently, we find that  the computational code only takes a few loops, or even one loop, to converge in low temperature region. And especially,  the approach has produced good agreements with  experimental and our theoretical results \cite{liupssb,LiuIan2}.

\vspace{0.5cm}
\begin{acknowledgements}
%{\footnotesize
Z.-S. Liu is supported   by National Natural Science Foundation of China
 under grant No.~11274177 and University of Macau,  H. Ian by the FDCT of Macau under grant 013/2013/A1,
University of Macau under grants MRG022/IH/2013/FST and
MYRG2014-00052-FST, and National
 Natural Science Foundation of China under Grant No.~11404415.
 %}
\end{acknowledgements}

\end{document}